\documentclass[twocolumn,showpaces,preprintnumbers,amsmath,amssymb]{revtex4}
\usepackage{amsmath}
\usepackage{graphicx}
\usepackage{color}

\begin{document}

\title{Electronic transport through graphene nanoribbons with
  Stone-Wales reconstruction at edges and interfaces}

\author{Jing Wang$^1$}
\author{Guiping Zhang$^{1}$}\email{bugubird_zhang@hotmail.com}
\author{Fei Ye$^2$}
\author{Xiaoqun Wang$^{1,3,4,5}$}\email{xiaoqunwang@ruc.edu.cn}
\affiliation{$^1$Department of Physics, Renmin University of China,
  Beijing 100872, China}
\affiliation{$^2$Department of Physics, South University of Science
  and Technology of China, Shenzhen 518055, China}
\affiliation{$^3$Beijing Laboratory of Optoelectronics Functional Materials and Micronano Device, Renmin University of China, Beijing 100872, China}
\affiliation{$^4$Department of Physics and Astronomy, Shanghai Jiao Tong University,
  Shanghai 200240, China}
\affiliation{$^5$ Collaborative Innovation Center of Advanced Microstructures, Nanjing 210093, China}
\date{\today}

\begin{abstract}

  In this paper, we study the conductance of the graphene
  nanoribbons(GNRs) in the presence of the Stone-Wales(S-W)
  reconstruction, using the transfer matrix method. The ribbon is
  connected with semi-infinite quantum wires as the leads. The S-W
  reconstruction occurs on the edges and the interfaces between the
  electrodes and ribbon. When the reconstruction occurs on the edges,
  the conductance is suppressed considerably if the gate voltage $V_g$
  takes intermediate values around $|V_{g}|\sim t_0$($t_0$ being the
  hopping amplitude of grahene) in both positive and negative energy
  regions. In contrast, if $V_g$ is close to the Dirac point or the band
  edges, the conductance is relatively insensitive to the edge
  reconstruction. The effect of edge reconstruction become less
  important with increasing ribbon width as expected.  The S-W
  reconstruction occurs also possibly at the interfaces. In this case,
  the reconstruction suppresses identically the conductance in the
  entire range of $V_g$ for armchair GNRs. For the zigzag GNRs, the
  conductance is strongly suppressed in the negative energy region,
  however the change of the conductance is relatively small in the
  positive energy region. We also analyze the transmission coefficients
  as functions of the channel index(the transverse momentum $k_y$ of the
  leads) for the neutral armchair GNRs with interface defects.
  Interestingly, there are two transmission peaks appearing at
  $k_{y}=\pi/3$ and $k_{y}=2\pi/3$ due to the unit cell doubling.

\end{abstract}

%\keywords:{strained graphene; transfer matrix method; band structure; topology}

\maketitle

\section{Introduction}

Graphene is one of the carbon allotropes with extraordinary electronic
properties\cite{graphene1,graphene2,graphene3-e}, that make it prominent
for potential applications in nano-electronic devices. In practice, one
need to take the edge effects into account, since they could change the
band structure and have important implications for the transport
properties of nanosize graphene. The conventional edges like armchair
and zigzag have been well studied, however they may not be robust
against edge reconstructions in some realistic
circumtances\cite{E-Edge-defect-1}.  Indeed, the reconstructed edges
have already been observed in experiments
\cite{E-Edge-defect-3,E-Edge-defect-4,E-Edge-defect-5}.

\begin{figure*}[htb]
  \centering
  \includegraphics[width=17cm]{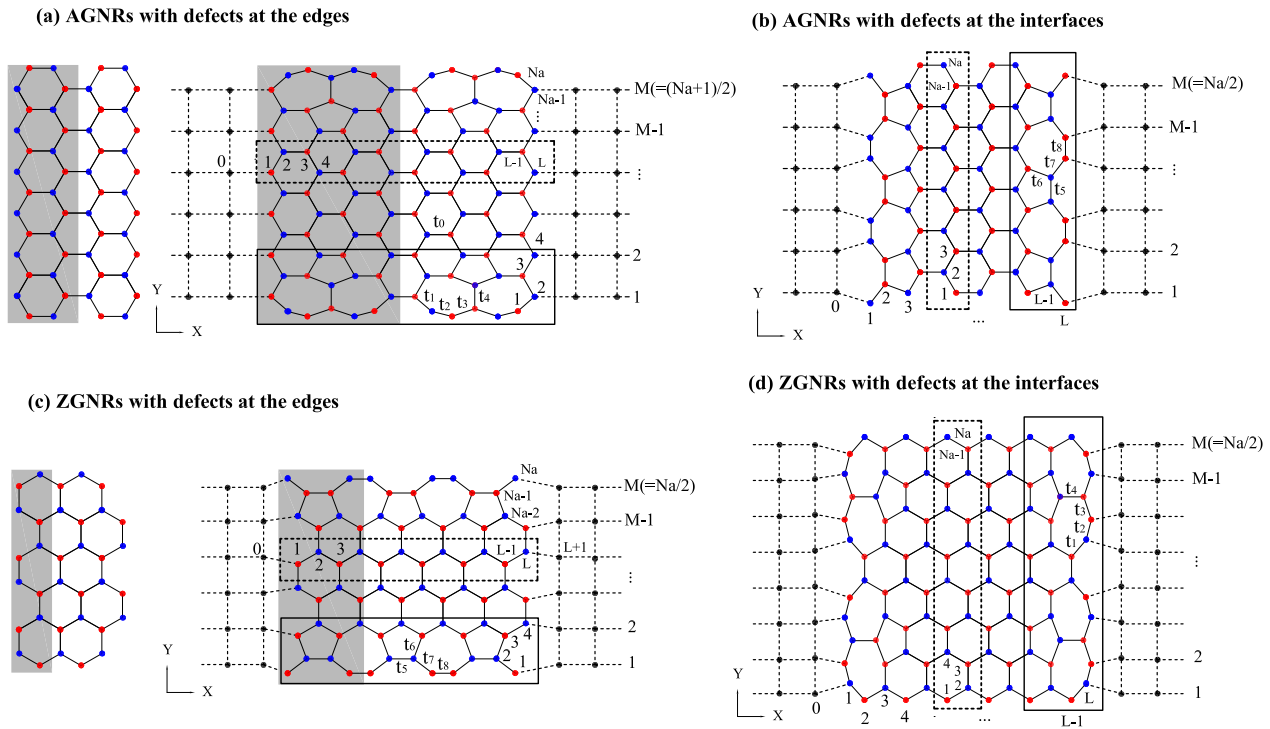}
  \caption{\label{QW_GRs} (Color online.)  The schematic illustration of
    S-W reconstruction on the ribbon edges (a, c), and at the interfaces
    (b, d) between the ribbon and the electrodes. The square lattices
    with width $M$ are the leads, and the current flows in the
    $x$-direction. $L$ and $N_a$ denotes the length and width(in terms
    of carbon atom number) for the GNRs, respectively.}
\end{figure*}

One of the reconstructions is caused by the Stone-Wales(S-W) mechanism,
which is obtained by $90^{\circ}$ rotation of the carbon-carbon bond
without changing total atom number. In Fig. \ref{QW_GRs} (a) and (c), we
show two typical continuous edge reconstructions by the S-W mechanism,
ac-757 in armchair graphene nanoribbons(AGNR) and zz-57 in zigzag
graphene nanoribbons(ZGNR), respectively.  Without hydrogen passivation,
a single ac-757 defect contains two adjacent heptagons and one pentagon,
while the zz-57 reconstruction consists of alternating heptagon and
pentagon on the edges \cite{E-Edge-defect-5,stability-1}. In contrast to
the graphene nanoribbons(GNRs) with the ideal edges, the more realistic
S-W edge reconstruction breaks electron-hole(e-h) symmetry and changes
the band structures substantially\cite{S-W2,S-W4}, of which the effects
on the transport of GNRs have attracted attension
recently\cite{S-W2,S-W1,S-W3,S-W4}.  It was found in Ref. \cite{S-W1}
that in the wide and short GNRs the conductance change due to the S-W
edge reconstruction is inconspicuous when the gate voltage is close to
the Dirac point. While, in another Ref.\cite{S-W2} it is shown that in
the narrow and long GNRs, the S-W defects at the edges modify the band
structure and result in strong backward scattering, that suppresses the
conductance. These two seemingly different senarios may be
attributed to the ribbon size, and also the distribution of the S-W
defects. In this paper we study the transport through GNRs with the S-W
defects periodically distributed on the ribbon edges.
% These two seemingly different senarios are
% integrated in the present transfer matrix study on GNRs with different
% sizes.
We also study the different effects of the S-W reconstructions
when they occur at different positions, i.e, the ribbon edges or the
heterojunctions between the ribbon and the electrodes.

The paper is organized as following. Our model and method are briefly
introduced in Sec. 2, and more details can be find in
Refs.\cite{tm1,tm2}. The effect of S-W reconstruction is given in Sec.3,
where we only consider the ac-757 and zz-57 reconstructions.  The
conclusion and summary are given in Sec.4.

\section{model and method}
The GNR is put in connection with semi-infinite quantum wires as the
electrodes which are simulated by the square lattices. Since the edge
reconstruction is inevitable in the fabrication and transfer of GNRs, it
can occur finally on both the ribbon edge and the interfaces between the
GNRs and electrodes. Figs. \ref{QW_GRs} shows the geometry of GNRs with
ac-757 and zz-57 reconstruction on the edges or interfaces. The
Hamiltonian defined on this network can be expressed in a tight binding
form
\begin{equation}
\label{eq:4}
\hat{H}=-\sum_{\langle ij,i^{'}j^{'}\rangle}t_{ij,i^{'}j^{'}} \hat{C}^{\dag}_{ij}\hat{C}_{i^{'}j^{'}}
-V_{g}\sum_{ij}\hat{C}^{\dag}_{ij}\hat{C}_{ij},
\end{equation}
where $V_{g}$ is the gate voltage applied on the GNRs and
$C^{\dag}_{ij}$ ($C_{ij}$) is the creation (annihilation) operator at
lattice site $i,j$.  $t_{ij,i^{'}j^{'}}$ is the hopping amplitude
between two nearest neighbors $ij$ and $i^{'}j^{'}$. The hopping
coefficients on the carbon hexagon edges is $t_0$ which is usually equal
to $2.7eV$, while those on the deformed bonds belonging to heptagons or
pentagons are taken from Refs. \cite{S-W1} as shown in Fig.~\ref{QW_GRs}
and Tab.~\ref{tab:1}. For simplicity, the hopping parameters in the
leads and the GNR-electrode interfaces are assumed to be $t_0$.  The
Fermi surface is set at zero energy, and the spin indices of electrons
are omitted simply for convenience.

\begin{table}[hpt]
  \centering
  \begin{tabular}[c]{|c|c|c|c|c|c|c|c|c|}
    \hline
    &  $t_1$ &    $t_2$ &    $t_3$ &    $t_4$ &    $t_5$ &    $t_6$ &
                   $t_7$ &    $t_8$  \\
    \hline%\hline
    $t_i/t_0$ & 0.982 & 1.189 & 0.991 & 1.018 & 0.982 & 0.957 & 1.009 & 1.189 \\
    \hline
  \end{tabular}
  \caption{The hopping amplitudes for the deformed bonds in unit of
    $t_0$, which are taken from Refs. \cite{S-W1}. $t_i$'s are defined in Fig.~\ref{QW_GRs}.}
  \label{tab:1}
\end{table}

In the leads with width $M$, the quantum states can be labeled by the
quantized transverse momenta $k_{y,n}=n\pi/(M+1)$ which correspond to
different conducting channels.  According to the Landauer-B\"{u}ttiker
formula, the conductance is the summation over all the transmission
coefficients $T_{n,n'}$ from channel $n$ to $n'$
\begin{equation}
\label{eq:8}
G=\frac{2e^{2}}{h}\sum_{n,n^{'}=1}^{M}T_{n,n'}.
\end{equation}
$T_{n,n'}$ can be calculated using the transfer matrix method
\cite{tm3,tm1,tm2,tm4,tm5}.

\begin{figure}[htb]
  \centering
  \includegraphics[width=8.5cm]{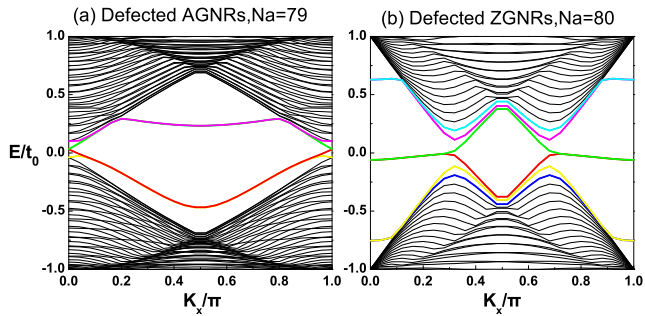}
  \caption{\label{band-structure} (Color online.) (a) is the band
    structure of AGNRs with ac-757 edge reconstruction, and (b) is that
    of ZGNRs with zz-57 edge reconstruction. Note that the unit cell is
    doubled by the reconstruction, and we take the length of the
    original unit cell in the pristine GNRs as unit, then the momentum
    $k_x$ belongs to the interval $[0,\pi)$.}
\end{figure}

We assume all the S-W defects align in a periodic pattern on the
boundaries of the GNRs with width $N_a$ and length $L$(see
Fig.~\ref{QW_GRs}(a, c)), which breaks the translational symmetry and the
unit cell is doubled. The randomly distributed defects are not discussed
in this article. The spectra of GNRs with periodic ac-757 and zz-57 edge
defects are shown in Fig.~\ref{band-structure} which are consistent with
those in Refs. \cite{S-W2,S-W4}. It turns out the e-h symmetry is
completely lost due to the existence of 5 and 7 atom rings on the edges.
As a consequence, for the AGNRs with ac-757 defects, the conduction and
valence bands touch at a positive energy $E>0$ for ${mod}(N_a,3)=1$(see
Fig.~\ref{band-structure}(a)). There appear four new subbands
corresponding to edge states \cite{S-W2}, indicated by colored lines in
Fig. \ref{band-structure} (a). For ZGNRs in the presence of the zz-57
reconstruction, the midgap band is no longer flat, but bends towards the
valence band(see Fig.~\ref{band-structure}(b)). Furthermore, there occur
additional midgap subbands with higher energy. These new features of the
energy spectra due to the edge reconstruction might affect the transport
properties finally.

\section{Transport through AGNRs and ZGNRs with S-W defects}
%\begin{figure*}[htb]
 % \centering
 % \includegraphics[width=18cm]{Band-structure-AGNRs.eps}
 % \caption{\label{BS-AGNRs-2} }
%\end{figure*}

\begin{figure}[htb]
  \centering
  \includegraphics[width=8.5cm]{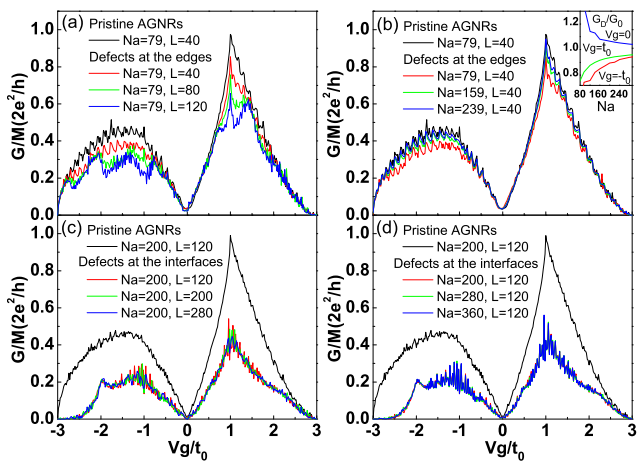}
  \caption{\label{AGNR-conductance}(Color online.) The normalized
    conductance $G/M$ as functions of the gate voltage $V_{g}$ for AGNRs
    with S-W reconstruction on the ribbon edges (a) and (b) and at the
    interfaces (c) and (d). The black curves are the conductance $G_0$
    of the pristine AGNRs for comparison. Inset of (b) shows the ratio
    $G_{D}/G_{0}$ as a function of the width at $V_{g}=0$ and
    $\pm t_0$.}
\end{figure}

We explore in details the transport properties of GNRs with ac-757 and
zz-57 reconstructions in this section.  Figures \ref{AGNR-conductance}
(a) and (b) show the normalized conductances $G_{D}/M$ of AGNRs with periodic S-W
defects on the ribbon edges as functions of the gate voltage $V_g$. The
overall e-h asymmetry of the conductance data is mainly due to the
odd-number-atom rings at the interfaces between the leads and the AGNRs
which makes the lattice non-bipartite\cite{strain-transport5}. The
conductance is suppressed considerably in the intermediate energy region
$0.7t_0<V_g<2t_0$ and $-2.7t_0<V_g<-0.7t_0$ by the ac-757 edge
reconstruction, where the trigonal warping effect is important which is
incompatible with the symmetry of heptagon and pentagon.  For $V_g$
close to zero energy or the band edges, $G_{D}$ is almost the same as
that of pristine AGNRs denoted by $G_{0}$, as shown in
Fig. \ref{AGNR-conductance} (a), except that the position of the minimal
conductance shifts slightly from zero towards a negative value. This is
due to the breaking e-h symmetry by the ac-757 edge reconstruction,
which results in the shift of the touch point of the valence and
conduction bands where the D.O.S. as well as the conductance is minimal.

In Figs. \ref{AGNR-conductance}(a) and (b), we also show the size effect
on the conductance. When the ribbon length increases, the electrons
experience more scattering from the S-W edge defects leading to the
further suppression of the conductance as shown in
Fig.~\ref{AGNR-conductance}(a). While, the increase of the ribbon width
has two effects, one is yielding more conducting channels leading to the
linear dependence of the conductance on the width, the other is
weakening the edge effect so that the conductance converges to that of
pristine AGNRs when the ribbon width is increased, as shown in
Fig.~\ref{AGNR-conductance}(b) and its inset.

When the zz-57 defects locate at the interfaces between the leads and the AGNR, the heptagons
and pentagons at the interfaces strongly scatter electrons and suppress
the conductance in the entire region of the gate voltages as shown
Figs.~\ref{AGNR-conductance}(c) and (d). In this case, we take a
relatively large ribbon width $N_a\ge 200$, so that we can focus on the
reconstruction at the interfaces and ignore those on the ribbon edges,
i.e., we are liberated to choose the conventional armchair edges. In
this setup, the conductance data are obviously insensitive to the size
as shown in Figs.~\ref{AGNR-conductance}(c) and (d).

\begin{figure}[htb]
  \centering
  \includegraphics[width=8.5cm]{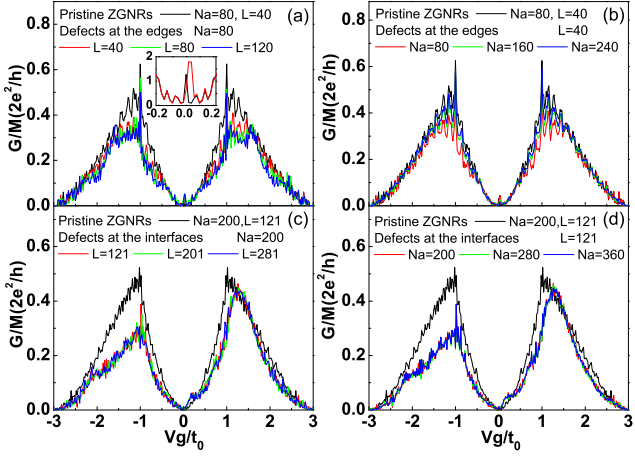}
  \caption{\label{ZGNR-conductance}(Color online.)  The normalized
    conductance $G/M$ as functions of the gate voltage $V_{g}$ for ZGNRs
    with S-W reconstruction on the ribbon edges (a) and (b) and at the
    interfaces (c) and (d). The black curves are the conductance $G_0$
    of the pristine ZGNRs for comparison. Inset of (a) shows the data of
    $G_{0}$ (black line) and $G_{D}$ (red line) with $L=40$ and $Na=80$
    in the energy region $[-0.2t_0,0.2t_0]$.  }
\end{figure}

Next, we consider the transport properties of ZGNRs with the S-W
reconstruction. Without edge reconstruction, the lattice of the ZGNR
connected with the leads is still bipartite unlike the case of
AGNRs\cite{strain-transport5}, therefore the conductance data remains
e-h symmetric as shown with the black curves in
Fig.~\ref{ZGNR-conductance}. When the zz-57 reconstruction is presented
on the edges, the e-h symmetry is broken due to the occurrence of
heptagons and pentagons, and the midgap band is not flat anymore(see
Fig.~\ref{band-structure}(b)). This is reflected in the conductance data
near $V_g=0$ in the inset of Fig.~\ref{ZGNR-conductance}(a), where the
peak position shifts from zero(black curve) toward a positive value(red
curve) and is enhanced due to the dispersion of the midgap band induced
by the edge reconstruction. Except for the shift of the midgap peak, the
zz-57 edge reconstruction barely changes the conductance data for small
gate voltage. This situation holds roughly in the energy interval
$|V_g|<0.5t_0$ and $|V_g|>2.3t_0$. For the intermediate energy region
$0.5t_0<|V_g|<2.3t_0$, the influence of the edge reconstruction is
obvious. Figure \ref{ZGNR-conductance}(b) shows the width dependence of
the conductance data, which also implies the edge effect is not
important in wide ribbons.

Figures \ref{ZGNR-conductance}(c) and (d) show the conductance data with
the ac-757 reconstruction at the interfaces between the leads and the
ZGNR. A relatively large width $N_a=200$ is taken so that the ribbon
edges do not have essential effect on the bulk transport, and can be
choosen as zigzag pattern for the simplicity of computation. The e-h
symmetry is obviously broken. The conductance in the negative energy
region is strongly suppressed by the scattering from the heptagons and
pentagons at the interfaces. In the positive energy region, the
influence of the ac-757 deformation is not strong, and besides the
reduction of the conductance as expected, there is a small region
$0<V_g<0.2t_0$ where the conductance is actually enhanced. Similar to
the case of AGNRs, the conductance is insensitive to the ribbon sizes
when the S-W reconstruction occurs at the interfaces.

\begin{figure}[htb]
  \centering
  \includegraphics[width=8.6cm]{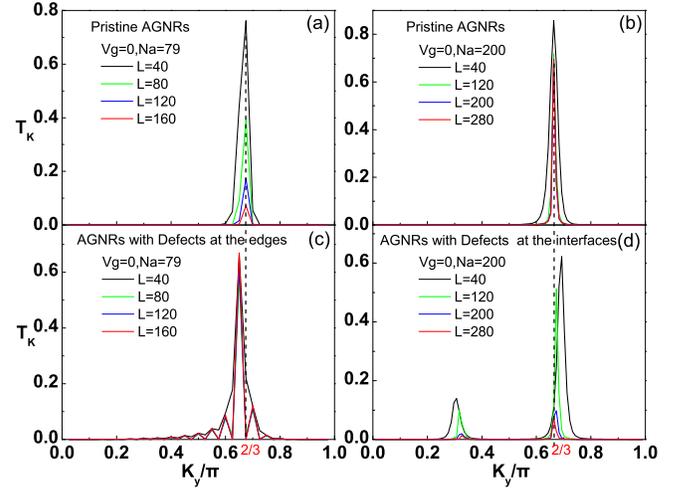}
  \caption{\label{TK-A}(Color online.)  The transmission coefficients in
    different $k_y$-channels for AGNRs without any reconstructions (a,
    b), with the S-W reconstruction on the edges (c) and at the
    interfaces (d). The vertical dashed lines mark the position of
    $2\pi/3$.}
\end{figure}

Finally, we show the transmission coefficients at $V_g=0$ through
different channels $k_{y,n}$, which is related to the states localized at
the interfaces of the electrodes and the
GNRs\cite{tm4,tm5,strain-transport5}. For pristine AGNRs connected with
the leads, it is the zigzag pattern that occurs at the interfaces. As
well known there are zero energy states localized at the zigzag
boundaries with momenta ranging from $2\pi/3$ to $4\pi/3$ \cite{Zigzag-EdgeState,Zigzag-EdgeState2}, which are
also involved in electron transmission as long as their localization
length is comparable with the ribbon length. This leads to the
transmission peak around
$k_y=2\pi/3$ as shown in
Figs.~\ref{TK-A}(a) and (b). The peak position is not sensitive to the
ribbon size, however the intensity is strongly dependent on the ribbon
length (see Fig.~\ref{TK-A}(a)) since increasing the ribbon length
reduces the number of the midgap states extending from one interface to
the other \cite{tm4}.

When the ac-757 reconstruction occurs at the edges of AGNRs, the e-h
symmetry is broken and the touch point of the conduction and valence
bands shifts upward as shown in Fig. \ref{band-structure} (a). As a
consequence, the maximal transmission peak moves slightly from
$k_{y}=2\pi/3$ toward a smaller value, accompanied by additional
satellite peaks as shown in Fig. \ref{TK-A}(c). These satellite peaks
arise because of the boundary condition of the zigzag interface is
modified by the zz-57 edge reconstruction.

\begin{figure}[htb]
  \centering
  \includegraphics[width=8.6cm]{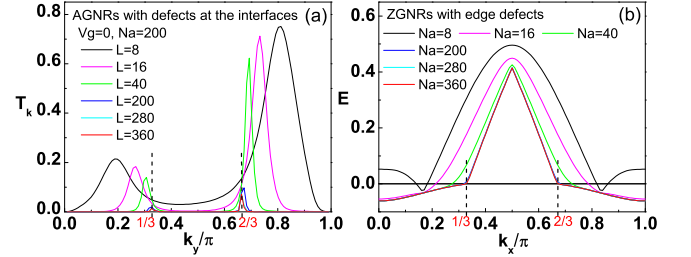}
  \caption{\label{TK-interface}(Color online.) (a) is the transmission
    coefficients in different $k_y$-channels for ANGRs with S-W
    reconstruction at the interfaces with various lengths $L$ . (b) is
    the subband structures relavant for the transport at zero gate
    voltage in ZGNRs with the S-W edge reconstruction(see also
    Fig.~\ref{band-structure} (b)). The black dashed lines mark the
    momenta $\pi/3$ and $2\pi/3$.}
\end{figure}

When the S-W reconstruction occur at the interfaces, there is an
additional zero energy transmission peak at $k_{y}=\pi/3$ with smaller
intensity, besides the original one at $k_y=2\pi/3$, as shown in
Fig. \ref{TK-A} (d).  The positions of these two peaks deviate from
$\pi/3$ and $2\pi/3$ for small ribbon length $L$, though they converge
to $\pi/3$ and $2\pi/3$ eventually as $L$ increases as shown in
Fig.~\ref{TK-interface}(a). As expected, this variation coincide with
that of the zero energy momenta with the width $N_a$ of the
corresponding ZGNRs with zz-57 edge reconstruction shown in
Fig.~\ref{TK-interface}(b). However these momenta given in
Fig.~\ref{TK-interface} don not have a naive one-to-one correspondence,
because the momentum $k_x$ in Fig.~\ref{TK-interface}(b) belongs to the
folded Brillouin zone(BZ), while $k_y$ in Fig.~\ref{TK-interface}(a)
corresponds to the unfolded BZ. In fact, for ZGNRs with zz-57 edge
reconstruction, the translational symmetry is broken with unit cell
doubled, then the two states with momenta $k$ and $\pi+k$ in the
unfolded BZ are mixed leading to new states with momentum $k$ in the
folded BZ. Therefore in Fig.~\ref{TK-interface} (b), the zero energy
state at $k_x=2\pi/3$ in the folded BZ corresponds to the mixing of two
momenta $2\pi/3$ and $5\pi/3$ in the unfolded BZ, which match the two
conducting channels with $k_y\sim 2\pi/3$ and $\pi/3$ in the leads,
respectively. The same holds for the zero energy state at $k_x=\pi/3$ in
the folded BZ, which corresponds to the mixing of the two momenta
$\pi/3$ and $4\pi/3$ in the unfolded BZ also corresponding to the two
conducting channels in the leads also with $k_y\sim\pi/3$ and $2\pi/3$.
Thus, we obtain two transmission peaks at $\pi/3$ and $2\pi/3$ in
Fig.~\ref{TK-interface} (a).

\section{Conclusion and summary}

In the presence of edge reconstructions, the band structure of GNRs
change substantially. For example, the e-h symmetry may be lost
completely and the midgap bands may appear in both AGNRs of ZGNRs with
their edges reconstructed by the S-W mechanism. This could eventually
affect the transport through GNRs, in particular those with small size,
which is crucial for designing the GNR-based nano-devices.

When the S-W reconstruction occurs on the ribbon edges, the conductance
$G_D$ in the intermediate energy region around $V_g\sim t_0$ is
suppressed considerably, in contrast, $G_D$ changes little for $V_g$
close to zero energy or band edges. When the S-W reconstruction occurs
at the interfaces between the GNRs and the electrodes, the conductance
is suppressed in the entire region of $V_g$ for AGNRs, while the
conductance of the ZGNRs shows strong e-h asymmetry, i.e., it is strongly
suppressed in the negative energy region, but change only a little in
the positive energy region. Therefore, to pursue high conductance of
GNRs, the defects should be avoided in heterojunctions between the
electrodes and the GNRs.

We also study the transmission coefficients in different channels of
AGNRs with possible S-W reconstructions when $V_g=0$. If the zigzag
pattern occurs at the interfaces, there is only one transmission peak
at $k_y\sim 2\pi/3$, with possible satellite peaks in the presence of
ac-757 defects on the ribbon edges. However in the presence of zz-57
reconstruction at the interfaces, two transmission peaks appears at
$\pi/3$ and $2\pi/3$ attributed to the unit cell doubled by the
alternating heptagon and pentagon. Our study may be useful in designing
the GNR-based nano-devices.

\section{Acknowledgements} This work is supported by the National Basic
Research Program of China (2012CB921704) and NSF of China (Grant
Nos. 11174363, 11204372, 11374135).


\begin{thebibliography}{99}

\bibitem{graphene1} K. S. Novoselov, A. K. Geim, S. V. Morozov, D. Jiang, M. I. Katsnelson, I. V. Grigorieva, S. V. Dubonos, and A. A. Firsov, Nature (London) 438, 197 (2005).
\bibitem{graphene2} K. S. Novoselov, A. K. Geim, S. V. Morozov, D. Jiang, Y. Zhang, S. V. Dubonos, I. V. Grigorieva, and A. A. Firsov, Science 306, 666 (2004).
\bibitem{graphene3-e} F. Miao, S. Wijeratne, Y. Zhang, U. C. Coskun, W. Bao, and C. N. Lau, Science 317, 1530 (2007).

\bibitem{E-Edge-defect-1} P. Koskinen, S. Malola, and H. H¡§akkinen, Phys. Rev. B 80, 073401
(2009).
%\bibitem{E-Edge-defect-2} S. Malola, H. H¡§akkinen, and P. Koskinen, Eur. Phys. J. D 52, 71
%(2009).

\bibitem{E-Edge-defect-3} C. Casiraghi, A. Hartschuh, H. Qian, S. Piscanec, C. Georgi, A. Fasoli, K. S. Novoselov, D. M. Basko, and A. C. Ferrari, Nano Lett. 9, 1433 (2009).

\bibitem{E-Edge-defect-4} C. \"{O} Girit, J. C. Meyer, R. Erni, M. D. Rossell, C. Kisielowski, Li Yang, Cheol-Hwan Park, M. F. Crommie, M. L. Cohen, S. G. Louie, and A. Zettl, Science 323, 1705 (2009).
\bibitem{E-Edge-defect-5} P. Koskinen, S. Malola, and H. Hakkinen, Phys. Rev. Lett. 101,
115502 (2008).

\bibitem{stability-1} B. Huang, M. Liu, N. Su, J.Wu,W. Duan, B. Gu, and F. Liu, Phys.
Rev. Lett. 102, 166404 (2009).

\bibitem{S-W4} J. N. B. Rodrigues, P. A. D. Gon\c{c}alves, N. F. G. Rodrigues, R. M. Ribeiro, J. M. B. Lopes dos Santos, and N. M. R. Peres, Phys. Rev. B 84, 155435 (2011).

\bibitem{S-W2} S. Ihnatsenka and G. Kirczenow, Phys. Rev. B 88, 125430 (2013).

\bibitem{S-W3} S. M.  -M. Dubois, A. L. Bezanilla, A. Cresti, F. Triozon, B. Biel, J. C. Charlier, and S. Roche, ACS nano 4, 1971 (2010).

\bibitem{S-W1} P. Hawkins, M. Begliarbekov, M. Zivkovic, S. Strauf, and C. P. Search, J. Phys. Chem. C 116, 18382 (2012).
\bibitem{tm1} S. J. Hu, W. Du, G. P. Zhang, M. Gao, Z. Y. Lu and X. Q. Wang, Chin. Phys. Lett. 29, 057201 (2012).
\bibitem{tm2} M. Gao, G. P. Zhang, Z. Y. Lu, Comput. Phys. Commun. 185, 856 (2014).
\bibitem{tm3} Y. Yin, S. J. Xiong, Phys. Lett. A 317, 507 (2003).
\bibitem{tm4} G. P. Zhang and Z. J. Qin, Chem. Phys. Lett. 516, 225 (2011).
\bibitem{tm5} G. P. Zhang, C. Z. Wang and K. M. Ho, Phys. Lett. A 375, 1043 (2011).
\bibitem{strain-transport5} J. Wang, G. P. Zhang, F. Ye and X. Q. Wang, arXiv:1411.1529; J. Phys.: Condens. Matt. (2015), in press.

%\bibitem{band structure-AGNR2} Y. W. Son, M. L. Cohen and S. G. Louie, Phys. Rev. Lett. 97, 216803 (2006).


\bibitem{Zigzag-EdgeState} K. Nakada, M. Fujita, G. Dresselhaus, and M. S. Dresselhaus, Phys. Rev. B 54, 17954 (1996).

\bibitem{Zigzag-EdgeState2} K. Wakabayashi, M. Fujita, H. Ajiki, and M. Sigrist, Phys. Rev. B 59, 8271 (1999).


\end{thebibliography}
\end{document}